\documentstyle[eqsecnum,prd,aps,epsf]{revtex}
\begin{document}
 \makeatletter
 \newdimen\ex@
 \ex@.2326ex
 \def\dddot#1{{\mathop{#1}\limits^{\vbox to-1.4\ex@{\kern-\tw@\ex@
  \hbox{\tenrm...}\vss}}}}
 \makeatother
\thispagestyle{empty}
{\baselineskip0pt
\leftline{\large\baselineskip16pt\sl\vbox to0pt{\hbox{\it Department of Physics}
               \hbox{\it Osaka City  University}\vss}}
\rightline{\large\baselineskip16pt\rm\vbox to20pt{\hbox{OCU-PHYS-175}
            \hbox{\today} 
\vss}}%
}
\vskip3cm
\begin{center}{\large 
{\bf Response of Interferometric Detectors 
to Scalar Gravitational Waves
}}
\end{center}
\begin{center}
 {\large Ken-ichi Nakao$^{1}$, Tomohiro Harada$^{2}$, Masaru Shibata$^{3,4}$, 
Seiji Kawamura$^{5}$ and Takashi Nakamura$^{6}$
} \\
{\em $^{1}$Department of Physics,~Osaka City University} \\
{\em Osaka 558-8585,~Japan}\\
{\em $^{2}$Department of Physics,~Waseda University} \\
{\em Oh-kubo, Shinjuku-ku, Tokyo 169-8555,~Japan} \\
{\em $^{3}$Department of Earth and Space Science, Graduate School of 
Science, Osaka University} \\
{\em Toyonaka, Osaka 560-0043, ~Japan}\\
{\em $^{4}$Department of Physics, University of Illinois at 
Urbana-Champaign, Urbana, IL 61801, USA} \\
{\em $^{5}$National Astronomical Observatory} \\
{\em Mitaka, Tokyo 181-8588,~Japan} \\
{\em $^{6}$Yukawa Institute of Theoretical Physics,~Kyoto University} \\
{\em Kyoto 606-8502,~Japan}
\end{center}
\begin{abstract}

We rigorously analyze 
the frequency response functions and antenna 
sensitivity patterns of three types of interferometric detectors 
to scalar mode of gravitational waves which is predicted to exist 
in the scalar-tensor theory of gravity. 
By a straightforward treatment,  
we show that the antenna sensitivity pattern of 
the simple Michelson interferometric detector depends 
strongly on the wave length $\lambda_{\rm SGW}$ 
of the scalar mode of gravitational waves if $\lambda_{\rm SGW}$ 
is comparable to the arm length of the interferometric detector. 
For the Delay-Line and Fabry-Perot 
interferometric detectors with arm length much shorter than 
$\lambda_{\rm SGW}$, however, the antenna sensitivity patterns depend 
weakly on $\lambda_{\rm SGW}$ even though $\lambda_{\rm SGW}$ is comparable 
to the effective path length of those interferometers. 
This agrees with the result obtained by Maggiore and Nicolis.


\end{abstract}

\section{Introduction}

To detect gravitational waves in the frequency band between $\sim 100$ and 
$\sim 1000$Hz, the TAMA300 has been constructed \cite{Ref:TAMA} 
and TAMA300 team has 
already carried out several data taking runs \cite{Tagoshi}. 
Other kilometer-size interferometric 
detectors such as LIGO, VIRGO and GEO600 
are also now under construction and will be in operation within a couple of 
years \cite{Ref:LIGO,Ref:VIRGO,Ref:GEO}.  In addition, 
the Laser Interferometer Space Antenna (LISA) is now planned 
to be constructed in the next decade \cite{LISA} 
to detect gravitational waves in the frequency band between 
$\sim 10^{-4}$ and $\sim 0.1$Hz. 
Main purposes of these projects are (i) the detection of 
gravitational waves and (ii) the observation of relativistic 
astrophysical phenomena in a strong gravitational field. 
The purpose (i) is related to a fundamental  
question on the correctness of general relativity because 
the detection of gravitational waves could give a strong
constraint about number of modes and propagation speed of gravitational 
waves \cite{will0}. 

As an alternative candidate of general relativity, 
the scalar-tensor theory of gravity \cite{Ref:Berg,Ref:Wago} has
recently received a renewed interest. 
One of the reasons for its revival is that a theoretical approach 
to unification of interactions in terms of superstring theories 
suggests that the theory of gravity should be the dilatonic one 
(a kind of the scalar-tensor theory) rather than 
general relativity, {\it i.e.}, the pure 
tensor theory \cite{Ref:DP}. 

In \cite{Ref:SNN}, Shibata, Nakao and Nakamura 
investigated the gravitational 
collapse of a spherically symmetric dust ball to be a black hole 
in the framework of the Brans-Dicke theory 
(see also \cite{Ref:HCNN} and \cite{Ref:NOVAK}). 
They found that the amplitude of scalar gravitational waves can be 
\begin{equation}
\sim 10^{-21}\biggl({M \over 10M_{\odot}}\biggr)
\biggl({3{\rm kpc} \over r}\biggr)
\biggl({5000 \over \omega}\biggr),
\end{equation}
or equivalently
\begin{equation}
\sim 10^{-21}\biggl({M \over 10^6M_{\odot}}\biggr)
\biggl({3000{\rm Mpc} \over r}\biggr)
\biggl({5000 \over \omega}\biggr),
\end{equation}
with the frequency $\sim 100(10M_{\odot}/M)=
10^{-3}(10^6M_{\odot}/M)$Hz which is in an appropriate frequency range 
for kilometer-size detectors and LISA, respectively. Here, 
$M$, $M_{\odot}$, $r$, and $\omega$ are the mass of the collapsed 
object, solar mass, distance from the source to the detectors, and 
the so-called Brans-Dicke parameter, respectively. 
The noise level of the advanced kilometer-size detectors and LISA will 
be $\sim 10^{-23}$ for the most sensitive frequency band. 
Thus, although the present experimental limit 
is stringent $\omega>3300$ \cite{Ref:OSHIETE}, 
the amplitude of scalar gravitational waves may be high 
enough to be detected by the interferometric detectors 
if stellar mass black holes are formed in our galaxy or if 
supermassive black holes are formed in the cosmological scale.

As pointed out in \cite{Ref:SNN} and reanalyzed by 
Maggiore and Nicolis \cite{Ref:MN}, 
the antenna sensitivity pattern of interferometric detectors 
to the scalar mode is different form that of the tensor mode. 
This implies that if four interferometric detectors which are coincidently 
in operation detects gravitational waves, it is possible to 
ask whether the scalar mode exists or not 
using the difference of the arrival time and 
the difference of the antenna pattern among four detectors \cite{Ref:TSK}. 
Also, the peak frequency of the scalar mode 
is likely different from that of the 
tensor mode \cite{SW,SSM}, 
suggesting that the two modes could be distinguished 
in the frequency domain even in the case of one interferometer. 
Thus, for exploration of the correctness of general relativity using 
interferometric detectors, 
it is important to clarify the antenna sensitivity pattern and 
frequency response function not 
only for the tensor mode but for the scalar mode of gravitational waves. 
{}From this motivation, analyses 
have been carried out in \cite{Ref:SNN}\cite{Ref:MN} so far, but 
they were too crude \cite{Ref:SNN} or approximate \cite{Ref:MN}. 
In this article, we carry out a rigorous reanalysis 
of the frequency response function and 
antenna sensitivity pattern of the interferometric detector 
to the scalar mode without any approximation: 
We directly solve the Maxwell equation 
for the propagation of a laser ray and 
the geodesic equation for motion of mirrors and a beam 
splitter in contrast to a previous work \cite{Ref:MN}. 
Our straightforward treatment enables us 
to investigate the dependence of the antenna sensitivity 
pattern on the wave length of the scalar mode of gravitational 
waves without any ambiguity. 

This article is organized as follows. In Sec. II, we briefly review 
the linearized equation for 
gravitational waves in the framework of the scalar-tensor theory 
of gravity.  In Sec. III, we solve the Maxwell field equation 
in the spacetime with the scalar mode of gravitational waves to 
derive a plane wave solution which describes the path of 
a laser ray in the interferometric detector. 
In Sec. IV, we present explicit solutions for timelike 
geodesics in the presence of scalar gravitational 
waves, since the mirrors and beam splitter in the 
interferometric detector move along timelike geodesics. 
We present the dependence of the frequency response function and 
antenna sensitivity pattern of Michelson 
interferometric detectors 
on the frequency of the scalar mode of gravitational waves in Sec. V, 
and of Delay-Line (DL) and Fabry-Perot (FP) interferometric 
detectors in Sec. VI. Sec. VII is devoted to a summary. 

We adopt the unit in which the speed of light is unity. 
We follow the convention of the metric and Riemann tensor 
in \cite{Ref:WALD}. Throughout this article, we adopt 
the abstract index notation; The Latin indices $a,...,d$ denote 
a type of tensor, Greek indices components of a tensor, and 
two Latin indices, $i$ and $j$, 
the spatial components of a tensor.

\section{Gravitational Waves in the Scalar-Tensor Theory}

The field equation for the tensor part in the scalar-tensor theory of 
gravity is written in the form \cite{Ref:Berg,Ref:Wago}
\begin{equation}
 G_{ab}
={8\pi\over\Phi}T_{ab} + {\omega(\Phi)\over\Phi^{2}}
 \Bigl[(\nabla_{a}\Phi)\nabla_{b}\Phi 
 -{1\over2}g_{ab}(\nabla^{c}\Phi)
 \nabla_{c}\Phi\Bigr]
 +{1\over\Phi}(\nabla_{a}\nabla_{b}\Phi
 -g_{ab}\nabla^{c}\nabla_{c}\Phi), 
 \label{eq:G-eq}
\end{equation}
where $G_{ab}$ and $\nabla_{a}$ are the Einstein tensor 
and covariant derivative with 
respect to $g_{ab}$, respectively.  The function, 
$\omega(\Phi)$, is the coupling function 
which determines the coupling strength between the gravitational 
scalar field, $\Phi$, and the spacetime geometry. 
$T_{ab}$ is the energy-momentum tensor of matter or radiation. 
The field equations for the scalar field $\Phi$ and matter are given by
\begin{equation}
\nabla^{c}\nabla_{c}\Phi = {1 \over 2\omega+3}
\Bigl[8\pi T^{c}_{c}-{d\omega\over d\Phi}
(\nabla^{c}\Phi)\nabla_{c}\Phi \Bigr],
\label{eq:S-eq}
\end{equation}
and 
\begin{equation}
\nabla_{a}T^{ab}=0. \label{eq:T-eq}
\end{equation}
In the Brans-Dicke theory, $\omega$ is a finite constant and 
in general relativity, $\omega=\infty$. 
Note that measurement of a gravitational lensing angle of a light ray 
from a distant quasar by the gravity of the Sun now constrains 
$|\omega|$ to be $>3300$ \cite{Ref:OSHIETE}. 

Since the interferometric detectors are located in the wave zone of 
gravitational waves, we need to consider the linearized, 
vacuum $(T_{ab}=0)$ field equations. 
We write the metric tensor and gravitational scalar field in the forms
\begin{eqnarray}
g_{ab}&=&\eta_{ab}+h_{ab}, \\
\Phi&=&\Phi_{0}(1-\delta\Phi),
\end{eqnarray}
where $\eta_{ab}$ is the metric tensor of the background 
Minkowski spacetime and $\Phi_{0}$ is the value of $\Phi$ 
in the infinity. 
We assume that magnitude of each component of 
the perturbations $h_{ab}$ and $\delta\Phi$ is much smaller than unity. 
The linearized equations 
for the metric tensor and gravitational scalar field are written as 
\begin{eqnarray}
-{1\over2}\partial^{c}\partial_{c}{\bar h}_{ab}
-{1\over2}\eta_{ab}\partial_{c}\partial_{d}{\bar h}^{cd}
+\partial_{c}\partial_{(a}{\bar h}^{c}_{b)}
=-\partial_{a}\partial_{b}\delta\Phi
+\eta_{ab}\partial^{c}\partial_{c}\delta\Phi,
\label{eq:L-eq}
\end{eqnarray}
and
\begin{equation}
\partial^{c}\partial_{c}\delta\Phi=0,
\end{equation}
where 
\begin{eqnarray}
h^{a}_{b}&\equiv&\eta^{ac}h_{cb}, \\
{\bar h}_{ab}&\equiv& h_{ab}-{1\over2}\eta_{ab}h^{c}_{c}, \\
\partial^{c}&\equiv&\eta^{cd}\partial_{d}.
\end{eqnarray}
Introducing a new symmetric tensor field ${\tilde h}_{ab}$ 
defined by 
\begin{equation}
{\tilde h}_{ab}\equiv {\bar h}_{ab}+\eta_{ab}\delta\Phi,\label{eq211}
\end{equation}
Eq. (\ref{eq:L-eq}) is rewritten to
\begin{equation}
-{1\over2}\partial^{c}\partial_{c}{\tilde h}_{ab}
-{1\over2}\eta_{ab}\partial_{c}\partial_{d}{\tilde h}^{cd}
+\partial_{c}\partial_{(a}{\tilde h}^{c}_{b)}
=0. \label{eq:EL-eq}
\end{equation}
Equation (\ref{eq:EL-eq}) is completely the same as the linearized 
Einstein equation for the trace-reversed metric perturbations 
\cite{Ref:WALD}. 
This enables us to 
impose the transverse-traceless gauge condition on 
${\tilde h}_{ab}$ as in general relativity as 
\begin{eqnarray}
\partial_{a}{\tilde h}_{b}^{a}&=&0,\\
{\tilde h}_{a0}&=&0.
\end{eqnarray}
Under this gauge condition, the line element in the 
wave zone is written in the following form, 
\begin{equation}
ds^{2}=(1+\delta\Phi)\left\{-dt^{2}+
(\delta_{ij}+{h}_{ij}^{\rm TT})d{\bar x}^{i}d{\bar x}^{j}
\right\},
\label{eq:line-element}
\end{equation}
where $h_{ij}^{\rm TT}\equiv {\tilde h}_{ij}$ 
is the transverse-traceless spatial tensor which satisfies
\begin{equation}
\partial^{c}\partial_{c}h^{\rm TT}_{ij}=0.
\end{equation}
{}From Eq. (\ref{eq:line-element}), $h^{\rm TT}_{ij}$ and 
$\delta\Phi$ are regarded as the tensor and scalar modes of 
gravitational waves, respectively, which constitute 
three independent modes of gravitational waves 
in the scalar-tensor theory.
Hereafter, we refer to the scalar mode as scalar gravitational waves 
(SGW). 
 
In the rest of this article, we assume that the amplitude of 
the tensor mode $h^{\rm TT}_{ij}$ is much smaller than that 
of SGW for simplicity. 
Such a situation is realized when the 
source of the gravitational radiation is almost spherically symmetric. 
We also assume 
\begin{equation}
\delta\Phi=\delta\Phi(t+{\bar z}),
\end{equation}
in the coordinate system (\ref{eq:line-element}). 

\section{Maxwell Field}

A trajectory of a laser ray is determined by solving 
the source-free Maxwell equation as 
\begin{equation}
g^{ac}\nabla_{a}F_{cb}=0, \label{eq:Maxwell-eq}
\end{equation}
where 
\begin{equation}
F_{ab}\equiv \nabla_{a}A_{b}-\nabla_{b}A_{a}
=\partial_{a}A_{b}-\partial_{b}A_{a}.
\end{equation}
{}From Eq. (\ref{eq:line-element}), 
we find that the line element takes conformally flat form 
in the coordinate system adopted here and 
in the absence of the tensor modes. 
Then, Eq. (\ref{eq:Maxwell-eq}) is rewritten to 
\begin{equation}
g^{ac}\nabla_{a}F_{cb}=
(1+\delta\Phi)^{-2}\eta^{ac}\partial_{a}F_{cb}
=(1+\delta\Phi)^{-2}(\partial^{a}\partial_{a}A_{b}
-\partial^{a}\partial_{b}A_{a})=0.
\end{equation}
Adopting the Lorentz gauge condition $\partial^{a}A_{a}=0$, 
we obtain the equation for the vector potential $A_{a}$ as
\begin{equation}
\partial^{a}\partial_{a}A_{b}=0. \label{eq:cf-Maxwell-eq}
\end{equation}
Note that Eq. (\ref{eq:cf-Maxwell-eq}) 
is of the same form as the equation in the Minkowski spacetime. 

In the geometrical approximation, a laser ray is regarded as 
a monochromatic plane wave in the absence of SGW, {\it i.e.}, 
\begin{equation}
A_{\mu}={\cal A}_{\mu}\exp
\left\{i\Omega_{\rm L}(t-{\vec c}_{\rm L}\cdot {\vec r}
)\right\}, \label{eq:Maxwell-sol}
\end{equation}
where ${\cal A}_{\mu}$, $\Omega_{\rm L}$, ${\vec c}_{\rm L}$ 
and ${\vec r}$ are the constant amplitude, 
constant angular frequency 
of the laser ray for the rest observer, 
constant direction cosine of the laser ray 
and position vector in the conformal 3D space, respectively. 
Eq. (\ref{eq:cf-Maxwell-eq}) implies that even in the presence of SGW, 
the components of the vector potential $A_{\mu}$ 
is completely the same as in Eq. (\ref{eq:Maxwell-sol}), and 
the trajectory of the laser ray is straight in the conformal 
3D space, {\it i.e.}, 
\begin{equation}
{d{\vec r}\over dt}={\vec c}_{\rm L}.
\end{equation}

{}From the above result, one might think that the laser interferometric 
detector could not detect SGW since SGW do not affect 
the Maxwell field described in the conformally flat coordinate system. 
However this is the misunderstanding. We should take into account that the 
laser ray is repeatedly reflected by mirrors which are perturbed in 
the presence of SGW. 
This non-trivial perturbed motion causes variations of 
the vector potential $A_{\mu}$ and consequently generates side-band 
waves which are just observed by interferometric detectors. 

\section{Timelike Geodesics}

The mirrors and beam splitter are suspended by wires so that 
they can freely move at frequencies much higher than the 
pendulum frequencies.  
For simplicity, we assume that the emitter of laser rays is 
also a free mass. 
Then, the trajectories of these parts of detectors 
are determined by solving the timelike geodesics 
in the spacetime described by (\ref{eq:line-element}). 

Introducing the retarded and advanced time coordinates $u$ and $v$,
\begin{equation} 
u=t-{\bar z}~~~~~{\rm and}~~~~~v=t+{\bar z},
\end{equation}
equations for a timelike geodesic are given by
\begin{eqnarray}
{d\over d\tau}\left\{(1+\delta\Phi){d{\bar x}\over d\tau}\right\}&=&0, 
\label{eq:time-1}\\
{d\over d\tau}\left\{(1+\delta\Phi){d{\bar y}\over d\tau}\right\}&=&0, 
\label{eq:time-2}\\
{d\over d\tau}\left\{(1+\delta\Phi){dv\over d\tau}\right\}&=&0,
\label{eq:time-3}
\end{eqnarray}
and
\begin{equation}
{d\over d\tau}\left\{(1+\delta\Phi){dv\over d\tau}\right\}
=-{d\delta\Phi\over dv}. \label{eq:time-4}
\end{equation}

Hereafter, we assume that the mirrors 
and beam splitter are at rest in the absence of SGW. In this case, 
we can set $\delta\Phi=0$ using freedom to scale the coordinates. 
Integrating Eqs. (\ref{eq:time-1})$-$(\ref{eq:time-4}), 
we obtain
\begin{equation}
{d{\bar x}\over d\tau}={d{\bar y}\over d\tau}=0, \label{eq:velocity-1}
\end{equation}
and 
\begin{equation}
{dv\over d\tau}=C_{v}(1-\delta\Phi),
\end{equation}
where $C_{v}$ is an integration constant. 
{}From the normalization of the geodesic tangent vector and 
the above equation, we derive 
\begin{equation}
{du\over d\tau}={1\over C_{v}}.
\end{equation}
Since $d{\bar z}/d\tau=0$ and $dt/d\tau=1$ 
for $\delta\Phi=0$, 
the integration constant $C_{v}$ is equal to 1. As a result, we find 
\begin{eqnarray}
{dt\over d\tau}&=&1-{1\over2}\delta\Phi(v), \label{eq:velocity-2}\\
{d{\bar z}\over d\tau}&=&-{1\over2}\delta\Phi(v).\label{eq:velocity-3}
\end{eqnarray}
By integrating Eqs. (\ref{eq:velocity-1})--(\ref{eq:velocity-3}), 
the trajectory of a free mass is derived as 
\begin{eqnarray}
{\bar x}&=&{\bar x}_{\rm i}, \label{eq:xb-trajectory}\\
{\bar y}&=&{\bar y}_{\rm i}, \label{eq:yb-trajectory} \\
{\bar z}&=&-{1\over2}\int^{\tau}\delta\Phi(v)d\tau
={\bar z}_{\rm i}-{1\over2}\int_{-\infty}^{t+{\bar z}_{\rm i}}\delta\Phi(v)dv,
\label{eq:zb-trajectory}
\end{eqnarray}
where ${\bar x}_{\rm i}$, ${\bar y}_{\rm i}$ and ${\bar z}_{\rm i}$ 
are integration constants which specify the location of 
the free mass in this coordinate system (\ref{eq:line-element}) 
for $t\rightarrow-\infty$, and we have used 
\begin{equation}
{d\tau\over dv}=1+\delta\Phi.
\end{equation}

Then, we derive a trajectory of a free mass for an arbitrary incoming 
direction of SGW, which is obtained by a spatial rotation of 
the coordinate system (\ref{eq:line-element}) as
\begin{equation}
\left(
\begin{array}{c}
x\\
y\\
z\\
\end{array}
\right)
=
\left(
\begin{array}{ccc}
\cos\theta\cos\varphi & -\sin\varphi & \sin\theta\cos\varphi \\
\cos\theta\sin\varphi &  \cos\varphi & \sin\theta\sin\varphi \\
-\sin\theta           &  0           & \cos\theta \\
\end{array}
\right)
\left(
\begin{array}{c}
{\bar x}\\
{\bar y}\\
{\bar z}\\
\end{array}
\right)
~~,
\end{equation}
where the direction cosine ${\vec c}_{\rm SGW}$ 
of SGW in the new coordinate system is 
\begin{equation}
{\vec c}_{\rm SGW}=(-\sin\theta\cos\varphi,-\sin\theta\sin\varphi,
-\cos\theta). 
\end{equation}
The explicit form of the trajectory of the free mass in the 
new coordinate system is written as
\begin{eqnarray}
x&=&x_{\rm i}-{1\over2}\sin\theta\cos\varphi
  \int_{-\infty}^{t+\Lambda}\delta\Phi(v)dv, \label{eq:x-trajectory} \\
y&=&y_{\rm i}-{1\over2}\sin\theta\sin\varphi
  \int_{-\infty}^{t+\Lambda}\delta\Phi(v)dv, \label{eq:y-trajectory} \\
z&=&z_{\rm i}-{1\over2}\cos\theta
  \int_{-\infty}^{t+\Lambda}\delta\Phi(v)dv, \label{eq:z-trajectory}
\end{eqnarray}
where $x_{\rm i}$, $y_{\rm i}$ and $z_{\rm i}$ are constants which 
specify the location of the free mass in this new coordinate system 
in the infinite past $t\rightarrow-\infty$, and 
\begin{equation}
\Lambda\equiv
x_{\rm i}\sin\theta\cos\varphi+
y_{\rm i}\sin\theta\sin\varphi+
z_{\rm i}\cos\theta.
\end{equation}

Since the spacetime metric 
takes conformally flat form in the absence of the tensor modes, we 
might think that 
SGW would be longitudinal wave. However, from careful investigation 
of the geodesic deviation equation, we find that the tidal force 
associated with SGW acts along the directions orthogonal to the 
propagation direction of SGW \cite{Ref:Wago}. 
{}From this point of view, SGW should be regarded as the transverse wave. 
It is also worthy to note that even if free masses are at rest in 
the present coordinate, the relative distances between them 
are not necessarily constant if SGW exist. Converse is also true. 


\section{Detection of SGW by a Simple Michelson Interferometer}

\begin{figure}
        \centerline{\epsfxsize 10cm \epsfysize 10cm \epsfbox{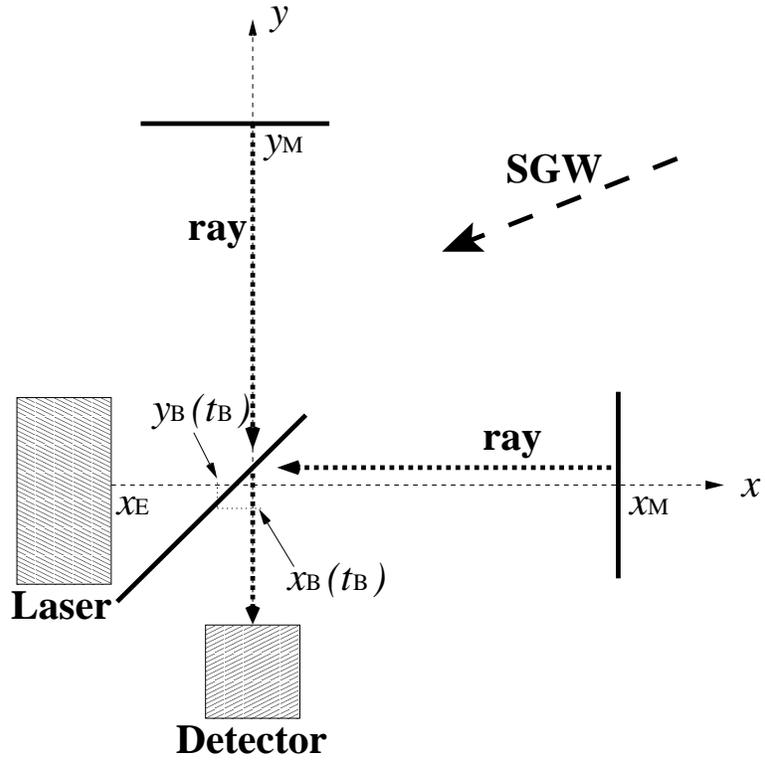}}
~(b)
        \caption{The definitions of various quantities for 
Michelson detectors. 
(a) A configuration in the absence of gravitational waves. (b)
A schematic configuration in the presence of 
scalar gravitational waves.}
\label{fg:M-detector}
\end{figure}

In the following, we identify the $x$ and $y$ axes 
with the directions of 
arms of the interferometric detector (cf, Fig. \ref{fg:M-detector}). 
We hereafter refer to the arm on the $x$ ($y$) axis as Arm1 (Arm2). 

In the Michelson interferometric 
detectors, a laser ray first reaches the beam splitter 
and then is divided into two rays; one enters into the Arm1 and 
the other into the Arm2. 
Each ray is reflected by the mirrors and then 
returns to the beam splitter again. 
The path difference
of the two arms is chosen in such a way that the interfering light is dark
at the anti-symmetric port for achieving the best signal-to-noise ratio. 
Here, we focus on two rays which finally reach $x=0$ on the beam splitter 
coincidently at $t=t_{\rm B}$ (cf, Fig. \ref{fg:M-detector}). 

To calculate the side-band waves of laser rays, 
we need the trajectories of the beam splitter and mirrors only in the 
$x$-$y$ plane. 
Using Eqs. (\ref{eq:x-trajectory})--(\ref{eq:z-trajectory}), 
we obtain the trajectory of the center of the beam splitter as
\begin{eqnarray}
x&=&x_{\rm B}(t)=-{1\over2}\sin\theta\cos\varphi 
\int_{-\infty}^{t}\delta\Phi(v)dv,\label{eq:Bx-trajectory}\\
y&=&y_{\rm B}(t)=-{1\over2}\sin\theta\sin\varphi 
\int_{-\infty}^{t}\delta\Phi(v)dv.
\label{eq:By-trajectory}
\end{eqnarray}
The trajectory 
of the mirror in Arm1 is given by
\begin{equation}
x=x_{\rm M}(t)=L_{1}-{1\over2}\sin\theta\cos\varphi
\int_{-\infty}^{t+X}\delta\Phi(v)dv,
\end{equation}
where
\begin{equation}
X=L_{1}\sin\theta\cos\varphi.
\end{equation}
Note that 
we do not need the $y$-coordinate of this mirror for calculating the
side-band waves since the motion in the $y$-direction 
does not affect the path difference in the linear order in $\delta \Phi$. 
{}From essentially the same procedure, 
the trajectory of the mirror in Arm2 is given by
\begin{equation}
y=y_{\rm M}(t)=L_{2}-{1\over2}\sin\theta\sin\varphi
\int_{-\infty}^{t+Y}\delta\Phi(v)dv,
\end{equation}
where
\begin{equation}
Y=L_{2}\sin\theta\sin\varphi.
\end{equation}
In this case, it is not necessary to know the $x$-coordinate of 
this mirror.

The trajectory of the emitter is written as
\begin{equation}
x=x_{\rm E}(t)=-\ell_{\rm E}-{1\over2}\sin\theta\cos\varphi
\int_{-\infty}^{t+X_{\rm E}}
\delta\Phi(v)dv, \label{eq:E-trajectory}
\end{equation}
where
\begin{equation}
X_{\rm E}=-\ell_{\rm E}\sin\theta\cos\varphi.
\end{equation}
Information of the $y$-coordinate of the emitter is also not 
necessary. 

Since we regard the emitter as a free mass, we need to carefully 
treat the phase of the laser ray. The amplitude $A$ of the 
laser ray at the emitter is written as
\begin{equation}
A={\cal A}_{\rm L}e^{i\Omega_{\rm L}\tau_{\rm E}},
\end{equation}
where ${\cal A}_{\rm L}$ and $\Omega_{\rm L}$ are constant and 
$\tau_{\rm E}$ is the proper time of the emitter. 
The relation between the coordinate time $t$ and  
proper time $\tau_{\rm E}$ is given by Eq. (\ref{eq:velocity-2}). 
Integrating this equation, we obtain
\begin{equation}
\tau_{\rm E}(t)
=t+{1\over2}\int_{-\infty}^{t+X_{\rm E}}\delta\Phi(v)dv.
\end{equation}
Thus, the amplitude $A$ is expressed as a function of $t$ as
\begin{equation}
A(t)={\cal A}_{\rm L}\exp\left\{i\Omega_{\rm L}\left(t+{1\over2}
\int_{-\infty}^{t+X_{\rm E}}\delta\Phi(v)dv
\right)\right\}. \label{eq:Amp-emitter}
\end{equation}

Then, we consider the ray entering into Arm1. 
Tracking from $t=t_{\rm B}$ toward the past, 
the laser ray reached the mirror in Arm1 at 
$t=t_{\rm B}-L_{1}+O(\delta\Phi)$, and then it reached the 
emitter at $t=t_{B}-2L_{1}-\ell_{\rm E}+O(\delta\Phi)$. 
As shown in Eq. (\ref{eq:Maxwell-sol}), the plane wave 
solution of the Maxwell equation is completely 
the same as that of the Minkowski spacetime 
in the conformally flat coordinate system. This implies that 
the amplitude $A_{1}$ of the laser ray at the beam splitter 
agrees with the amplitude $A$ at the emitter 
at $t=t_{\rm B}-r_{1}$, where
$r_{1}$ is the coordinate path length of this ray from the emitter 
to the final beam splitter. 
Using Eq. (\ref{eq:Amp-emitter}), we 
obtain the amplitude $A_{1}$ of the laser ray at the beam splitter 
on the output port side at $t=t_{\rm B}$ as 
\begin{equation}
A_{1}=A(t_{\rm B}-r_{1})=\rho_{\rm B}\nu_{\rm B} 
{\cal A}_{\rm L}\exp\left\{i\Omega_{\rm L}\left(t_{B}-r_{1}
+{1\over2}
\int_{-\infty}^{t_{\rm B}-r_{1}+X_{\rm E}}\delta\Phi(v)dv
\right)\right\}, \label{eq:Amp-1}
\end{equation}
where $\rho_{\rm B}$ is a reflection coefficient from the 
detector side ($-\rho_{\rm B}$ from the laser side) 
and $\nu_{\rm B}$ is transmission coefficient 
of the beam splitter.

From the mirror at $t=t_{\rm B}-L_{1}+O(\delta\Phi)$
to the beam splitter at $t=t_{\rm B}$, the coordinate path length 
of the laser ray is equal to $x_{\rm M}(t_{\rm B}-L_{1})+O(\delta\Phi^2)$, 
and  from the beam splitter at $t=t_{\rm B}-2L_{1}+O(\delta\Phi)$  
to the mirror at $t=t_{\rm B}-L_{1}+O(\delta\Phi)$, 
the coordinate path length is 
$x_{\rm M}(t_{\rm B}-L_{1})+O(\delta\Phi^2)$. 
{}From the emitter at $t=t_{\rm B}-2L_{1}-\ell_{\rm E}+O(\delta\Phi)$ 
to the beam splitter at $t=t_{\rm B}-2L_{1}+O(\delta\Phi)$, 
the coordinate path length is 
$-x_{\rm E}(t_{\rm B}-2L_{1}-\ell_{\rm E})+O(\delta\Phi^2)$. 
Thus, $r_{1}$ is given by
\begin{eqnarray}
r_{1}&=&x_{\rm M}(t_{\rm B}-L_{1})+x_{\rm M}(t_{\rm B}-L_{1})
-x_{\rm E}(t_{\rm B}-2L_{1}-\ell_{\rm E})
\nonumber \\
&=&2L_{1}+\ell_{\rm E}+{1\over2}
\left(
2\int_{-\infty}^{t_{\rm B}-L_{1}+X}
-\int_{-\infty}^{t_{\rm B}-2L_{1}-\ell_{\rm E}+X_{\rm E}}\right)
\delta\Phi(v)dv.
\end{eqnarray}
Substituting the above expression into Eq. (\ref{eq:Amp-1}), 
the amplitude $A_{1}$ is written in the form 
\begin{equation}
A_{1}=\rho_{\rm B}\nu_{\rm B}
{\cal A}_{\rm L}e^{i\Omega_{\rm L}(t_{\rm B}-2L_{1}-\ell_{\rm E})}
\left\{1+\Delta A_{1}+O(\delta\Phi^{2})\right\},
\end{equation}
where
\begin{eqnarray}
\Delta A_{1}&\equiv&
{i\over2}\Omega_{\rm L}\Biggl\{
\sin\theta\cos\varphi\left(2\int_{-\infty}^{t_{\rm B}-L_{1}+X}
-\int_{-\infty}^{t_{\rm B}-2L_{1}-\ell_{\rm E}+X_{\rm E}}\right)
+\int_{-\infty}^{t_{\rm B}-2L_{1}-\ell_{\rm E}+X_{\rm E}}
\Biggr\}\delta\Phi(v)dv.
\end{eqnarray}

{}From the same procedure as that for 
the laser ray entering into Arm1,  
the amplitude $A_{2}$ at the beam splitter 
on the output port side for the laser ray entering into Arm2 
is expressed in the form 
\begin{equation}
A_{2}=A(t_{\rm B}-r_{2})=-\rho_{\rm B}\nu_{\rm B} 
{\cal A}_{\rm L}\exp\left\{i\Omega_{\rm L}\left(t_{B}-r_{2}
+{1\over2}
\int_{-\infty}^{t_{\rm B}-2L_{2}-\ell_{\rm E}+X_{\rm E}}\delta\Phi(v)dv
\right)\right\}, \label{eq:Amp-2}
\end{equation}
where $r_{2}$ is the coordinate path length 
of this ray from the emitter to the final beam splitter. 

As shown in Fig. \ref{fg:M-detector}, 
the trajectory of the laser ray considered here 
is restricted on the straight line $x=0$ when it goes through 
the Arm2. Thus, this laser ray comes back to the beam splitter 
first at 
\begin{equation}
y=y_{\rm B}(t_{\rm B}-2L_{2})-x_{\rm B}(t_{\rm B}-2L_{2}), 
\end{equation}
and finally at 
\begin{equation}
y=y_{\rm B}(t_{\rm B})-x_{\rm B}(t_{\rm B}). 
\end{equation}
{}From the same procedure as in the case of Arm1, 
the coordinate path length $r_{2}$ from the emitter 
to the final beam splitter is given by
\begin{eqnarray}
r_{2}&=&y_{\rm M}(t_{\rm B}-L_{2})
-\left\{y_{\rm B}(t_{B})-x_{\rm B}(t_{B})\right\} \nonumber \\ 
&+&y_{\rm M}(t_{\rm B}-L_{2})-\left\{y_{\rm B}(t_{\rm B}-2L_{2})
-x_{\rm B}(t_{\rm B}-2L_{2})\right\} \nonumber \\
&-&x_{\rm E}(t_{\rm B}-2L_{2}-\ell_{\rm E}) \nonumber \\
&=&2L_{2}+\ell_{\rm E}
-{1\over2}\sin\theta\Biggl\{\cos\varphi
\left(\int_{-\infty}^{t_{\rm B}}+\int_{-\infty}^{t_{\rm B}-2L_{2}}
-\int_{-\infty}^{t_{\rm B}-2L_{2}-\ell_{\rm E}+X_{\rm E}}
\right) \nonumber \\
&+&\sin\varphi\left(\int_{t_{\rm B}}^{t_{\rm B}-L_{2}+Y}
+\int_{t_{\rm B}-2L_{2}}^{t_{\rm B}-L_{2}+Y}\right)
\Biggr\}\delta\Phi(v)dv.
\end{eqnarray}
Substituting the above equation into Eq. (\ref{eq:Amp-2}), 
we obtain
\begin{equation}
A_{2}=-\rho_{\rm B}\nu_{\rm B}
A_{\rm L}e^{i\Omega_{\rm L}(t_{\rm B}-2L_{2}-\ell_{\rm E})}
\left\{1+\Delta A_{2}+O(\delta\Phi^{2})\right\},
\end{equation}
at the final beam splitter, where
\begin{eqnarray}
\Delta A_{2}&\equiv&
{i\over2}\Omega_{\rm L}\Biggl\{
\sin\theta\cos\varphi\left(\int_{-\infty}^{t_{\rm B}}
+\int_{-\infty}^{t_{\rm B}-2L_{2}}
-\int_{-\infty}^{t_{\rm B}-2L_{2}-\ell_{\rm E}+X_{\rm E}}\right) \nonumber \\
&+&\sin\theta\sin\varphi\left(\int_{t_{\rm B}}^{t_{\rm B}-L_{2}+Y}
+\int_{t_{\rm B}-2L_{2}}^{t_{\rm B}-L_{2}+Y}\right)
+\int_{-\infty}^{t_{\rm B}-L_{2}-\ell_{\rm E}+X_{\rm E}}
\Biggr\}\delta\Phi(v)dv.
\end{eqnarray}

To maintain the output port to be dark in the absence of SGW 
$\delta\Phi=0$, 
the following condition should be satisfied 
\begin{equation}
\Omega_{\rm L}(L_{1}-L_{2})=m \pi,
\end{equation}
where $m$ is integer. 
Therefore, in the presence of SGW, 
the side-band waves are generated with the amplitude 
\begin{eqnarray}
A_{\rm out}&=&A_{1}+A_{2} \nonumber \\
&=&\rho_{\rm B}\nu_{\rm B} A_{\rm L}e^{i\Omega_{\rm L}
(t_{\rm B}-2L_{1}-\ell_{\rm E})}
(\Delta A_{1}-\Delta A_{2}) \nonumber \\
&=&\rho_{\rm B}\nu_{\rm B} A_{\rm L}\Omega_{\rm L}e^{i\Omega_{\rm L}
(t_{\rm B}-2L_{1}-\ell_{\rm E})}
\int_{-\infty}^{+\infty}\delta{\tilde\Phi}(\omega)e^{i\omega t_{\rm B}}
H_{\rm M}(\omega)d\omega,
\end{eqnarray}
where 
\begin{equation}
\delta{\tilde\Phi}(\omega)\equiv
{1\over 2\pi}\int_{-\infty}^{+\infty}e^{-i\omega v}\delta\Phi(v)dv,
\end{equation}
and $H_{\rm M}(\omega)$ is a frequency response function to SGW of  
Michelson interferometric detectors. (Note that hereafter $\omega$ does 
not denote the Brans-Dicke parameter but the angular frequency.)  
It is worthy to note that difference between 
the coordinate time $t_{\rm B}$ and the proper time at the detector  
is $O(\delta\Phi)$ so that it only contributes 
to a higher order correction for the side-band waves. 

The explicit form of $H_{\rm M}(\omega)$ is 
\begin{eqnarray}
H_{\rm M}(\omega)&=&
{1\over2\omega}\sin\theta\Biggl\{
\cos\varphi\left(2e^{i\omega L_{1}(\sin\theta\cos\varphi-1)}-1
-e^{-2i\omega L_{1}}\right) \nonumber \\
&-&\sin\varphi \left(2e^{i\omega L_{2}(\sin\theta\sin\varphi-1)}
-1-e^{-2i\omega L_{2}}\right)\Biggr\}. 
\end{eqnarray}

\begin{figure}
        \centerline{\epsfxsize 13cm \epsfysize 10cm \epsfbox{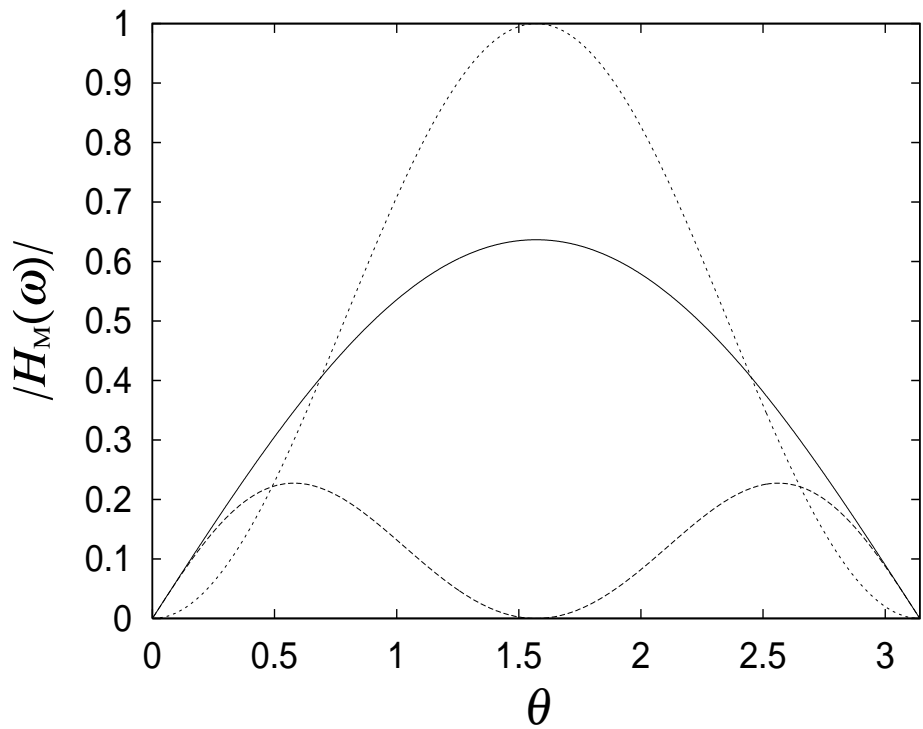}}
~(b)
        \caption{(a) The frequency response functions  
of Michelson interferometric detectors for
$\omega=\omega_{\rm M}$, 
$\omega_{\rm M}/2$ and $\omega_{\rm M}/100$ 
on $\theta=\pi/2$ as a function of $\varphi$ 
(the solid, dashed and dotted lines, respectively). 
 (b) The frequency response function of the same detector for the 
same $\omega$ as (a) 
on $\varphi=0$ is depicted as a function of $\theta$. 
In both cases, we take the units in which 
$L_{1}$ and $L_{2}$ are unity.}
\label{fg:pattern1}
\end{figure}

For simplicity, we restrict our attention to the case
\begin{equation}
L_{1}=L_{2}=L.
\end{equation}
Then, in the limit $\omega\rightarrow 0$, the response 
function becomes
\begin{equation}
\lim_{\omega\rightarrow 0}H_{\rm M}(\omega)=iL\sin^{2}\theta\cos2\varphi,
\label{eq:M-limit-value}
\end{equation}
which agrees with the antenna sensitivity 
pattern obtained by Maggiore and Nicolis \cite{Ref:MN}. 
However, for SGW 
of general frequency $\omega$, the antenna sensitivity 
pattern is different from Eq. (\ref{eq:M-limit-value})
(cf, Fig. \ref{fg:pattern1}). 

To see the dependence of the amplitude of 
$H_{\rm M}(\omega)$ on the 
frequency $\omega$ of SGW, we pay particular attention to 
the incident direction $\theta=\pi/2$ and $\varphi=0$. Then, we 
obtain 
\begin{equation}
|H_{\rm M}(\omega)|^{2}={\sin^{2}(\omega L)\over \omega^{2}}.
\label{eq:M-sensitivity}
\end{equation}
Thus, Michelson interferometric detectors 
are sensitive to SGW of $\omega \alt \omega_{\rm M}$ where
\begin{equation}
\omega_{\rm M}\equiv \pi/(2L).
\end{equation}

{}From Eq. (\ref{eq:M-limit-value}), we find that 
the longer the arm length $L$ is, the better the sensitivity 
of Michelson detectors is. 
However, longer arm length leads to smaller upper bound $\omega_{\rm M}$
of the sensitive frequency range. To optimize the 
laser interferometric detector against SGW of an aimed frequency, 
the arm length of the laser ray should be equal 
to a quarter of the wave length of SGW. 

The sensitivity patterns on $\theta=\pi/2$ plane 
for the frequency $\omega=\omega_{\rm M}$ of aimed SGW 
are shown in Fig. \ref{fg:pattern1}(a). 
Obviously, the sensitivity patterns for $\omega=\omega_{\rm M}$  and 
$\omega=\omega_{\rm M}/2$ are different 
from Eq. (\ref{eq:M-limit-value}). 
On the other hand,  
for $\omega=\omega_{\rm M}/100$, {\it i.e.}, 
when the wave length of SGW is $400L$, 
the sensitivity pattern is almost the same as 
that in Eq. (\ref{eq:M-limit-value}). 
These consequences imply that 
the sensitivity pattern of Michelson interferometric detectors 
depends strongly on the 
frequency in the neighborhood of $\omega=\omega_{\rm M}$.

\section{Detection of SGW by Delay-Line and Fabry-Perot Interferometers}

As explained in the previous section, 
the arm length of Michelson interferometric detectors should be equal  
to a quarter of the wave length of SGW to optimize the sensitivity. 
The wave length of SGW generated by a spherical gravitational collapse 
to a black hole, however, 
could be $\sim 100(M/M_{\odot})$km for stellar mass 
black holes \cite{Ref:SNN}. 
It is in practice difficult to construct 
a laser interferometric detector of such long arm length 
on the ground. Due to this reason, 
DL or FP interferometric detectors in which 
it is possible to make the effective 
optical path length much longer than the arm length have been 
adopted as ground-based detectors in practical projects. 
In this section, we study the frequency response function and 
antenna sensitivity pattern of DL and FP laser 
interferometric detectors. 

\begin{figure}
        \centerline{\epsfxsize 10cm \epsfysize 10cm \epsfbox{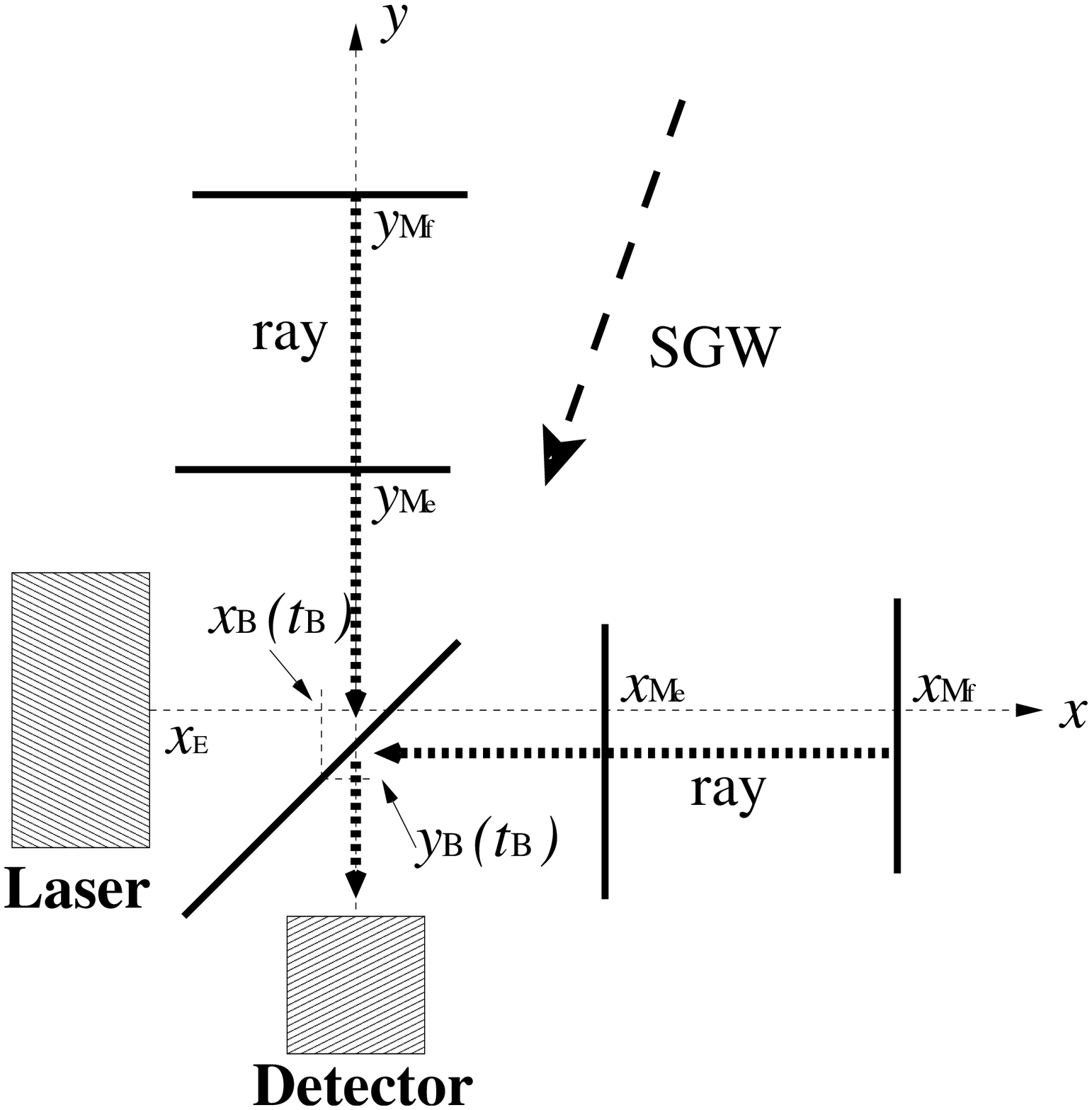}}
~(b)
        \caption{The definitions of various quantities for 
DL and FP interferometric detectors. 
(a) A configuration in the absence of gravitational waves. (b)
A schematic configuration in the presence of 
scalar gravitational waves.}
\label{fg:DF-detector}
\end{figure}

Following the notation in Sec. V, 
we identify the $x$ and $y$ axes 
with the direction of the arms of the interferometric detector 
(cf, Fig. \ref{fg:DF-detector}). 
We again refer to the arm on the $x$ ($y$) axis as Arm1 (Arm2). 
As in the case of Michelson detectors, 
we need the trajectories of the mirrors and beam splitter only  in the 
$x$-$y$ plane. 
The trajectory of the center of the beam splitter is given 
by Eqs. (\ref{eq:Bx-trajectory}) and (\ref{eq:By-trajectory}).  
The trajectories of the front mirror and of 
the end mirror in Arm1 are given by
\begin{eqnarray}
&&x=x_{\rm Mf}(t)=\ell_{1}-{1\over2}\sin\theta\cos\varphi
\int_{-\infty}^{t+X_{\rm f}}\delta\Phi(v)dv, \\
&&x=x_{\rm Me}(t)=L_{1}+\ell_{1}-{1\over2}\sin\theta\cos\varphi
\int_{-\infty}^{t+X_{\rm e}}\delta\Phi(v)dv,
\end{eqnarray}
where
\begin{eqnarray}
X_{\rm f}&=&\ell_{1}\sin\theta\cos\varphi, \\
X_{\rm e}&=&(L_{1}+\ell_{1})\sin\theta\cos\varphi.
\end{eqnarray}
The trajectories of the front mirror 
and the end mirror in Arm2 are given by
\begin{eqnarray}
&& y=y_{\rm Mf}(t)=\ell_{2}-{1\over2}\sin\theta\sin\varphi
\int_{-\infty}^{t+Y_{\rm f}}\delta\Phi(v)dv, \\
&& y=y_{\rm Me}(t)=L_{2}+\ell_{2}-{1\over2}\sin\theta\sin\varphi
\int_{-\infty}^{t+Y_{\rm e}}\delta\Phi(v)dv,
\end{eqnarray}
where
\begin{eqnarray}
Y_{\rm f}&=&\ell_{2}\sin\theta\sin\varphi, \\
Y_{\rm e}&=&(L_{2}+\ell_{2})\sin\theta\sin\varphi.
\end{eqnarray}
We do not need to calculate 
the $y$-coordinates of mirrors in Arm  1 
and the $x$-coordinates of 
mirrors in Arm2 to obtain side-band waves 
since the motion of these mirrors in these directions  
affects the path difference from the second order in $\delta \Phi$. 
The trajectory of the emitter and 
the amplitude $A$ of the laser ray at the emitter 
are given by Eq. (\ref{eq:E-trajectory}) 
and by Eq. (\ref{eq:Amp-emitter}), respectively, as in Michelson 
detectors. However, 
the definitions of $L_{1}$ and $L_{2}$ are different 
from those in Michelson detectors (cf, Fig. \ref{fg:DF-detector}).

\subsection{Delay-Line Interferometer}

As in Michelson interferometric detectors, a 
laser ray in DL detectors first reaches the beam splitter 
and then is divided into two rays; one enters into the Arm1 and 
the other into the Arm2. In the DL detectors, however, 
the ray entering into the regions between 
the front and end mirrors is reflected by these mirrors by ($2N-1$) times 
until returning to the beam splitter again
to increase an optical path length by $2N$ times 
as long as the arm length. This mechanism changes the response 
of the detector. 

As in Sec. IV, 
we pay attention to two rays which finally reach the beam splitter 
coincidently at $t=t_{\rm B}$ (cf, Fig. \ref{fg:DF-detector}).  
First, we consider the ray entering into Arm1.
For notational convenience, we introduce
\begin{eqnarray}
t_{{\rm f}(n)}&=&t_{\rm B}-\left\{\ell_{1}+(2n-2)L_{1}\right\}, \\
t_{{\rm e}(n)}&=&t_{\rm B}-\left\{\ell_{1}+(2n-1)L_{1}\right\}, \\
t_{{\rm E}(n)}&=&t_{\rm B}-\left(2\ell_{1}+\ell_{\rm E}+2nL_{1}\right).
\end{eqnarray}
{}From the same procedure as in Michelson interferometric detectors, 
the amplitude $A_{1}$ of the $N$-fold ray  
at the beam splitter on the output port side 
is expressed in the form 
\begin{equation}
A_{1}=\rho_{\rm B}\nu_{\rm B} 
A_{\rm L}\exp\left\{i\Omega_{\rm L}\left(t_{B}-r_{(N)}
+{1\over2}
\int_{-\infty}^{t_{{\rm E}(N)}+X_{\rm E}}\delta\Phi(v)dv
\right)\right\}, \label{eq:Amp-N}
\end{equation}
where the definition of $\rho_{\rm B}$ and $\nu_{\rm B}$ are the same 
as for the Michelson detector,  
and $r_{(N)}$ is the coordinate path length 
of this ray from the emitter to the final beam splitter. 

In the following calculation for the coordinate 
path length $r_{(N)}$, we assume that 
the trajectory of the ray is parallel to the $x$-axis. 
Strictly speaking, the ray is not parallel to the $x$-axis 
in a real DL interferometric detector, but the effect of it on the 
path length is the second order in $\delta \Phi$. 
Then we obtain
\begin{eqnarray}
r_{(N)}&=&x_{{\rm Mf}}(t_{{\rm f}(1)})
+\sum_{k=1}^{N}\left\{2x_{{\rm Me}}(t_{{\rm e}(k)})
-x_{{\rm Mf}}(t_{{\rm f}(k)})-x_{{\rm Mf}}(t_{{\rm f}(k+1)})
\right\}
+x_{{\rm Mf}}(t_{{\rm f}(N+1)})
-x_{{\rm E}}(t_{{\rm E}(N)})
\nonumber \\
&=&2\ell_{1}+\ell_{\rm E}+2NL_{1}+\Delta R_{(N)},
\label{eq:path1}
\end{eqnarray}
where
\begin{eqnarray}
\Delta R_{(N)}&=&
-{1\over2}\sin\theta\cos\varphi
\Biggl\{
\int_{-\infty}^{t_{{\rm f}(1)}+X_{\rm f}}
+\int_{-\infty}^{t_{{\rm f}(N+1)}+X_{\rm f}}
-\int_{-\infty}^{t_{{\rm E}(N)}+X_{\rm E}}
+\sum_{k=1}^{N}\left(
\int_{t_{{\rm f}(k)}+X_{\rm f}}^{t_{{\rm e}(k)}+X_{\rm e}}
+\int_{t_{{\rm f}(k+1)}+X_{\rm f}}^{t_{{\rm e}(k)}+X_{\rm e}}\right)
\Biggr\}\delta\Phi(v)dv.
\end{eqnarray}
Substituting Eq. (\ref{eq:path1}) into Eq. (\ref{eq:Amp-N}), 
the amplitude $A_{1}$ becomes
\begin{equation}
A_{1}=\rho_{\rm B}\nu_{\rm B}
A_{\rm L}e^{i\Omega_{\rm L}(t_{\rm B}-2\ell_{1}-\ell_{\rm E}-2NL_{1})}
\left\{1+\Delta A_{1}+O(\delta\Phi^{2})\right\},
\end{equation}
where
\begin{equation}
\Delta A_{1}\equiv-i\Omega_{\rm L}\left(\Delta R_{(N)}
-{1\over2}\int_{-\infty}^{t_{{\rm E}(N)}+X_{\rm E}}\delta\Phi(v)dv
\right).
\end{equation}

Next we consider the ray entering into Arm2. 
As above, we introduce
\begin{eqnarray}
{\bar t}_{{\rm f}(n)}&=&t_{\rm B}-\left\{\ell_{2}+(2n-2)L_{2}\right\}, \\
{\bar t}_{{\rm e}(n)}&=&t_{\rm B}-\left\{\ell_{2}+(2n-1)L_{2}\right\}, \\
{\bar t}_{{\rm B}(n)}&=&t_{\rm B}-(2\ell_{2}+2nL_{2}), \\
{\bar t}_{{\rm E}(n)}&=&t_{\rm B}-(2\ell_{2}+\ell_{\rm E}+2nL_{2}),
\end{eqnarray}
and in addition, 
\begin{equation}
w_{\rm B}(t)\equiv y_{\rm B}(t)-x_{\rm B}(t).
\end{equation}
The path length ${\bar r}_{(N)}$ of the $N$-fold ray 
from the emitter to the final beam splitter is given by
\begin{eqnarray}
{\bar r}_{(N)}&=&y_{{\rm Mf}}({\bar t}_{{\rm f}(1)})-w_{\rm B}(t_{\rm B})
+\sum_{k=1}^{N}\left\{2y_{{\rm Me}}({\bar t}_{{\rm e}(k)})
-y_{{\rm Mf}}({\bar t}_{{\rm f}(k)})
-y_{{\rm Mf}}({\bar t}_{{\rm f}(k+1)})\right\}
+y_{{\rm Mf}}({\bar t}_{{\rm f}(N+1)})
-w_{\rm B}({\bar t}_{{\rm B}(N)})
-x_{\rm E}({\bar t}_{{\rm E}(N)})
\nonumber \\
&=&2\ell_{2}+\ell_{\rm E}+2NL_{2}+\Delta {\bar R}_{(N)},
\end{eqnarray}
where
\begin{eqnarray}
\Delta {\bar R}_{(N)}&=&-{1\over2}\sin\theta\sin\varphi 
\Biggl\{\int_{t_{\rm B}}^{{\bar t}_{\rm f(N+1)}+Y_{\rm f}}
+\int_{{\bar t}_{{\rm B}(N)}}^{{\bar t}_{\rm f(1)}+Y_{\rm f}}
+\sum_{k=1}^{N}\left(
\int_{{\bar t}_{{\rm f}(k)}+Y_{\rm f}}^{{\bar t}_{{\rm e}(k)}+Y_{\rm e}}
+\int_{{\bar t}_{{\rm f}(k+1)}+Y_{\rm f}}^{{\bar t}_{{\rm e}(k)}+Y_{\rm e}}
\right)
\Biggr\}\delta\Phi(v)dv \nonumber \\
&+&{1\over2}\sin\theta\cos\varphi
\biggl(
\int_{-\infty}^{{\bar t}_{{\rm E}(N)}+X_{\rm E}}
-\int_{-\infty}^{t_{\rm B}}
-\int_{-\infty}^{{\bar t}_{{\rm B}(N)}} 
\biggr)\delta\Phi(v)dv.
\end{eqnarray}

{}From the same procedure as in the case of Arm1, 
the amplitude $A_{2}$ of the laser ray entering into 
Arm2 is 
\begin{equation}
A_{2}=-\rho_{\rm B}\nu_{\rm B}
A_{\rm L}e^{i\Omega_{\rm L}(t_{\rm B}-2\ell_{2}-\ell_{\rm E}-2NL_{2})}
\left\{1+\Delta A_{2}+O(\delta\Phi^{2})\right\},
\end{equation}
at the final beam splitter on the output port side, where
\begin{equation}
\Delta A_{2}\equiv-i\Omega_{\rm L}\left(\Delta {\bar R}_{(N)}
-{1\over2}\int_{-\infty}^{{\bar t}_{{\rm E}(N)}+X_{\rm E}}\delta\Phi(v)dv
\right).
\end{equation}

To maintain the output port to be dark in the absence of SGW, 
the following condition should be satisfied 
\begin{equation}
\Omega_{\rm L}(\ell_{1}+NL_{1})=
\Omega_{\rm L}(\ell_{2}+NL_{2})+m \pi,
\end{equation}
where $m$ is integer. 
Thus, in the presence of SGW, 
side-band waves are generated with the amplitude 
\begin{eqnarray}
A_{\rm out}&=&A_{1}+A_{2} \nonumber \\
&=&\rho_{\rm B}\nu_{\rm B} A_{\rm L}e^{i\Omega_{\rm L}
(t_{\rm B}-2\ell_{1}-\ell_{\rm E}-2NL_{1})}
(\Delta A_{1}-\Delta A_{2}) \nonumber \\
&=&\rho_{\rm B}\nu_{\rm B} A_{\rm L}\Omega_{\rm L}e^{i\Omega_{\rm L}
(t_{\rm B}-2\ell_{1}-\ell_{\rm E}-2NL_{1})}
\int_{-\infty}^{+\infty}\delta{\tilde\Phi}(\omega)e^{i\omega t_{\rm B}}
H_{\rm DL}(\omega)d\omega,
\end{eqnarray}
where $H_{\rm DL}(\omega)$ is a frequency response function to SGW of  
DL interferometric detector. 
The explicit form of $H_{\rm DL}(\omega)$ is 
\begin{eqnarray}
H_{\rm DL}(\omega)&=&
{1\over2\omega}\Biggl[
\sin\theta\cos\varphi\biggl\{e^{i\omega(X_{\rm f}-\ell_{1}-2NL_{1})}-1
+e^{i\omega(X_{\rm f}-\ell_{1})}-e^{-2i\omega(\ell_{2}+NL_{2})} \nonumber \\
&+&e^{i\omega(X_{\rm E}-\ell_{\rm E})}
(e^{-2i\omega(\ell_{2}+NL_{2})}-e^{-2i\omega(\ell_{1}+NL_{1})}) \nonumber \\
&+&{e^{-2i\omega L_{1}}(1-e^{-2i\omega NL_{1}})\over 1-e^{-2i\omega L_{1}}}
(2e^{i\omega(X_{\rm e}-\ell_{1}+L_{1})}
-e^{i\omega(X_{\rm f}-\ell_{1}+2L_{1})}
-e^{i\omega(X_{\rm f}-\ell_{1})})\biggr\} \nonumber \\
&-&\sin\theta\sin\varphi\biggl\{e^{i\omega(Y_{\rm f}-\ell_{2}-2NL_{2})}-1
+e^{i\omega(Y_{\rm f}-\ell_{2})}-e^{-2i\omega(\ell_{2}+NL_{2})} \nonumber \\
&+&{e^{-2i\omega L_{2}}(1-e^{-2i\omega NL_{2}})\over 1-e^{-2i\omega L_{2}}}
(2e^{i\omega(Y_{\rm e}-\ell_{2}+L_{2})}
-e^{i\omega(Y_{\rm f}-\ell_{2}+2L_{2})}
-e^{i\omega(Y_{\rm f}-\ell_{2})})\biggr\} \nonumber \\
&-&e^{i\omega(X_{\rm E}-\ell_{\rm E})}
(e^{-2i\omega(\ell_{2}+NL_{2})}-e^{-2i\omega(\ell_{1}+NL_{1})})
\Biggr] \label{eq:DL-Rfunction}
\end{eqnarray}

For simplicity, we restrict our attention to the case
\begin{equation}
\ell_{1}=\ell_{2}=\ell~~~~{\rm and}~~~~
L_{1}=L_{2}=L.
\end{equation}
In the practical DL interferometric detectors, 
$\ell \ll L$ and $\omega L < 1$ for the 
most sensitive frequency band of the detector. 
Then, for $\omega\ell \ll 1$, 
the frequency response function of DL 
detector becomes
\begin{eqnarray}
H_{\rm DL}(\omega)&\simeq &
{\sin(\omega NL)\over2\omega\sin(\omega L)}
e^{-i\omega(N-1)L}
\sin\theta\biggl\{
\cos\varphi(2e^{i\omega L(\sin\theta\cos\varphi-1)}
-1-e^{-2i\omega L}) \nonumber \\
&-&\sin\varphi(2e^{i\omega L(\sin\theta\sin\varphi-1)}
-1-e^{-2i\omega L})
\biggr\}\nonumber \\
&=&{\sin(\omega NL)\over\sin(\omega L)}
e^{-i\omega(N-1)L}H_{\rm M}(\omega;L),
\end{eqnarray}
where $H_{\rm M}(\omega;L)$ is the frequency 
response function of Michelson interferometric detector 
of the arm length $L_{1}=L_{2}=L$. 
Hence, the antenna sensitivity pattern of DL interferometric 
detectors is the same as that for Michelson detectors 
of the arm length $L$, except for the overall factor 
depending on $\omega$. 
In the limit $\omega\rightarrow 0$, the frequency 
response function for two types of detectors coincides  as 
\begin{equation}
\lim_{\omega\rightarrow0}H_{\rm DL}(\omega)
=iNL\sin^{2}\theta\cos2\varphi.
\label{eq:DL-limit-value}
\end{equation}
However, for finite $\omega$, this is not the case. 
To see the dependence of the amplitude of $H_{\rm DL}$ on the 
frequency $\omega$ of SGW, we pay particular attention to the case 
that the incident direction of SGW is $\theta=\pi/2$ and $\varphi=0$. 
In this case, we find 
\begin{equation}
|H_{\rm DL}|^{2}={\sin^{2}(N\omega L)\over \omega^{2}}.
\end{equation}
This implies that the sensitivity of 
DL interferometric detector 
of $N$-fold is high for $\omega\alt \omega_{\rm DL}$, where
\begin{equation}
\omega_{\rm DL}\equiv \pi/(2NL).
\end{equation}
To optimize DL interferometric detectors 
against SGW of an aimed wave length $\lambda_{\rm SGW}$, 
the folding number $N$ is usually set to be 
$\lambda_{\rm SGW}/4L$ so that 
$\omega_{\rm DL}=2\pi/\lambda_{\rm SGW}$ is satisfied. 
As mentioned above, in the case of practical interferometric detectors 
constructed on the ground, $N$ is much larger than unity. 
Since the dependence of the antenna sensitivity pattern 
on the frequency of SGW is the same as that of Michelson one, 
and since the DL detector of $N\gg1$ is sensitive 
for the wave length $\lambda_{\rm SGW}>2\pi/\omega_{\rm DL}\gg L$, 
the antenna sensitivity pattern of DL for aimed SGW is 
almost the same as Eq. (\ref{eq:DL-limit-value}), 
even if the wave length of SGW is comparable to the optical path 
length. In other words, we can consider that the amplitude of 
SGW is effectively constant 
in computing the response of the detectors.

\subsection{Fabry-Perot Interferometer}

The FP interferometric detectors form the Fabry-Perot cavity between 
the front and end mirrors. We define that 
the reflection coefficients of the front mirrors for the 
light ray incident from the FP cavity side and laser side are 
$\rho_{\rm M}$ and $-\rho_{\rm M}$, respectively, 
and that the transmission coefficients of these are 
$\nu_{\rm M}$. Hereafter, we consider a loss-free system in which 
\begin{equation}
\nu_M^2 + \rho_M^2=1. 
\end{equation}
We assume that the end mirrors 
have a perfect reflectivity, {\it i.e.}, the reflection coefficients 
of those are $-1$. On the other hand, the reflection and 
transmission coefficients of the beam splitter 
are assumed to be $\rho_{\rm B}$ and $\nu_{\rm B}$ as before.

As in the case of Michelson and DL interferometric detectors, 
we focus on laser rays which finally reach $x=0$ on the beam splitter 
at $t=t_{B}$. First, we consider the ray entering into Arm1. 
The path length $r_{(0)}$ of the ray which does not enter into 
the FP cavity is
\begin{eqnarray}
r_{(0)}&=&2x_{{\rm Mf}}(t_{{\rm f}(1)})-x_{{\rm E}}(t_{{\rm E}(0)}) 
\nonumber \\
&=&2\ell_{1}+\ell_{\rm E}-{1\over2}\sin\theta\cos\varphi
\left(
2\int_{-\infty}^{t_{{\rm f}(1)}+X_{\rm f}}
-\int_{-\infty}^{t_{{\rm E}(0)}+X_{\rm E}}\right)\delta\Phi(v)dv.
\end{eqnarray}
{}From the same procedure as in the case of DL interferometer, 
the path length $r_{(n)}$ of the ray which enters into the FP cavity and 
is reflected by the front and end mirrors by $(2n+1)$ times until 
returning to the beam splitter is 
given by
\begin{equation}
r_{(n)}=r_{(0)}+2nL_{1}+\Delta r_{(n)},
\end{equation}
where
\begin{eqnarray}
\Delta r_{(n)}&=&-{1\over2}\sin\theta\cos\varphi\Biggl\{
\int_{t_{{\rm f}(1)}+X_{\rm f}}^{t_{{\rm f}(n+1)}+X_{\rm f}}
-\int_{t_{{\rm E}(0)}+X_{\rm E}}^{t_{{\rm E}(n)}+X_{\rm E}}
+\sum_{k=1}^{n}\left(
\int_{t_{{\rm f}(k)}+X_{\rm f}}^{t_{{\rm e}(k)}+X_{\rm e}}
+\int_{t_{{\rm f}(k+1)}+X_{\rm f}}^{t_{{\rm e}(k)}+X_{\rm e}}\right)
\Biggr\}\delta\Phi(v)dv.
\end{eqnarray}
As in the derivation of Eq. (\ref{eq:Amp-N}), 
the amplitude $A_{1(0)}$ of the path length $r_{(0)}$ 
at the beam splitter on the side of the output port is obtained as
\begin{equation}
A_{1(0)}=-\nu_{\rm B}\rho_{\rm B}\rho_{\rm M}
e^{i\Omega_{\rm L}(t_{\rm B}-2\ell_{1}-\ell_{\rm E}-\delta r_{(0)})},
\end{equation}
where
\begin{equation}
\delta r_{(0)}\equiv -{1\over 2}\left\{
2\sin\theta\cos\varphi \int_{-\infty}^{t_{{\rm f}(1)}+X_{\rm f}}
+(1-\sin\theta\cos\varphi) \int_{-\infty}^{t_{{\rm E}(0)}
+X_{\rm E}}
\right\}
\delta\Phi(v)dv.
\end{equation}
The amplitude $A_{1(n)}$ of the path length $r_{(n)}$ at the 
beam splitter on the side of the output port is given by
\begin{equation}
A_{1(n)}=-\left({\nu_{\rm M}\over\rho_{\rm M}}\right)^{2}A_{1(0)}
(-\rho_{\rm M}e^{-2i\Omega_{\rm L}L_{1}})^{n}
e^{-i\Omega_{\rm L}\delta r_{(n)}}, 
\end{equation}
where
\begin{equation}
\delta r_{(n)}=\Delta r_{(n)}-{1\over2}
\int_{t_{{\rm E}(0)}+X_{\rm E}}^{t_{{\rm E}(n)}+X_{\rm E}}
\delta\Phi(v)dv.
\end{equation}
Hence when the ray reaches the beam splitter on the side of the 
output port at $t=t_{\rm B}$, 
the amplitude $A_{1}$ of the laser ray is
\begin{equation}
A_{1}=\sum_{n=0}^{\infty}A_{1(n)}
=-\nu_{\rm B}\rho_{\rm B}\rho_{\rm M}A_{\rm L}
e^{i\Omega_{\rm L}(t_{\rm B}-2\ell_{1}-\ell_{\rm E})}
(a_{1}+\Delta A_{1})+O(\delta\Phi^{2}),
\end{equation}
where 
\begin{eqnarray}
a_{1}&\equiv& 1+{\nu_{\rm M}^{2}e^{-2i\Omega_{\rm L}L_{1}}\over
\rho_{\rm M}(1+\rho_{\rm M}e^{-2i\Omega_{\rm L}L_{1}})},\\
\Delta A_{1}&\equiv&-i\Omega_{\rm L}
\left\{a_{1}\delta r_{(0)}
-\left({\nu_{\rm M}\over \rho_{\rm M}}\right)^{2}
\sum_{n=1}^{\infty}(-\rho_{\rm M}e^{-2i\Omega_{\rm L}L_{1}})^{n}
\delta r_{(n)}\right\}.
\end{eqnarray}

Next, we consider the ray entering into Arm2. 
{}From the same procedure as that for the Arm 1, 
we obtain the path lengths ${\bar r}_{(0)}$ and ${\bar r}_{(n)}$ as 
\begin{eqnarray}
{\bar r}_{(0)}&=&2y_{{\rm Mf}}({\bar t}_{{\rm f}(1)})
-w_{\rm B}(t_{\rm B})
-w_{\rm B}(t_{\rm B}-2\ell_{2})
-x_{{\rm E}}({\bar t}_{{\rm E}(0)})
\nonumber \\
&=&2\ell_{2}+\ell_{\rm E}-{1\over2}\sin\theta
\Biggl\{
\sin\varphi
\left(
\int_{t_{\rm B}}^{{\bar t}_{\rm f(1)}+Y_{\rm f}}
+\int_{t_{\rm B}-2\ell_{2}}^{{\bar t}_{\rm f(1)}+Y_{\rm f}}\right)
+\cos\varphi\left(
\int_{-\infty}^{t_{\rm B}}+\int_{-\infty}^{t_{\rm B}-2\ell_{2}}
-\int_{-\infty}^{{\bar t}_{{\rm E}(0)}+X_{\rm E}}
\right)\Biggr\}
\delta\Phi(v)dv.
\end{eqnarray}
\begin{equation}
{\bar r}_{(n)}={\bar r}_{(0)}+2nL_{2}+\Delta {\bar r}_{(n)},
\end{equation}
where
\begin{eqnarray}
\Delta {\bar r}_{(n)}&=&-{1\over2}\sin\theta\sin\varphi
\Biggl\{
\int_{{\bar t}_{{\rm f}(1)}+Y_{\rm f}}^{{\bar t}_{{\rm f}(n+1)}+Y_{\rm f}}
+\int_{{\bar t}_{{\rm B}(n)}}^{t_{\rm B}-2\ell_{2}}
+\sum_{k=1}^{n}\left(
\int_{{\bar t}_{{\rm f}(k)}+Y_{\rm f}}^{{\bar t}_{{\rm e}(k)}+Y_{\rm e}}
+\int_{{\bar t}_{{\rm f}(k+1)}+Y_{\rm f}}^{{\bar t}_{{\rm e}(k)}+Y_{\rm e}}
\right)
\Biggr\}\delta\Phi(v)dv \nonumber \\
&+&{1\over2}\sin\theta\cos\varphi
\left(\int_{{\bar t}_{{\rm B}(n)}}^{t_{\rm B}-2\ell_{2}}
+\int_{{\bar t}_{{\rm E}(0)}+X_{\rm E}}^{{\bar t}_{{\rm E}(n)}+X_{\rm E}}
\right)\delta\Phi(v)dv.
\end{eqnarray}
Here, the path length ${\bar r}_{(n)}$ of the ray 
is computed for the ray which enters into the FP 
cavity and is reflected by the front and end mirrors by $(2n-1)$ 
times until returning to the beam splitter.

The amplitude $A_{2}$ of the laser ray entering into 
Arm2 is written in the form
\begin{equation}
A_{2}=\nu_{\rm B} \rho_{\rm B}\rho_{\rm M}
e^{i\Omega_{\rm L}(t_{\rm B}-2\ell_{2}-\ell_{\rm E})}
(a_{2}+\Delta A_{2})+O(\delta\Phi^{2}),
\end{equation}
where we assume that 
the ray reaches the beam splitter on the side of the output port 
at $t=t_{\rm B}$, and 
\begin{eqnarray}
a_{2}&\equiv& 1+{\nu_{\rm M}^{2}e^{-2i\Omega_{\rm L}L_{2}}\over
\rho_{\rm M}(1+\rho_{\rm M}e^{-2i\Omega_{\rm L}L_{2}})}, \\
\Delta A_{2}&\equiv&-i\Omega_{\rm L}
\left\{a_{2}\delta {\bar r}_{(0)}
-\left({\nu_{\rm M}\over \rho_{\rm M}}\right)^{2}
\sum_{n=1}^{\infty}(-\rho_{\rm M}e^{-2i\Omega_{\rm L}L_{2}})^{n}
\delta {\bar r}_{(n)}\right\},\\
\delta {\bar r}_{(0)}&\equiv& 
-{1\over2}\Biggl\{\sin\theta\sin\varphi\left(
\int_{t_{\rm B}}^{{\bar t}_{\rm f(1)}+Y_{\rm f}}
+\int_{t_{\rm B}-2\ell_{2}}^{{\bar t}_{\rm f(1)}+Y_{\rm f}}
\right)
+\sin\theta\cos\varphi \left(
\int_{-\infty}^{t_{\rm B}}+\int_{-\infty}^{t_{\rm B}-2\ell_{2}}\right)
\nonumber \\
&+&(1-\sin\theta\cos\varphi)\int_{-\infty}^{{\bar t}_{\rm E(0)}
+X_{\rm E}}\Biggr\}
\delta\Phi(v)dv, \\
\delta {\bar r}_{(n)}&\equiv&\Delta {\bar r}_{(n)}-{1\over2}
\int_{{\bar t}_{{\rm E}(0)}+X_{\rm E}}^{{\bar t}_{{\rm E}(n)}+X_{\rm E}}
\delta\Phi(v)dv.
\end{eqnarray}

We assume that 
the following resonance condition holds in the absence of SGW, 
\begin{equation}
2\Omega_{\rm L}L_{i}=(2n_{i}+1)\pi,
\end{equation}
where $n_{i}$ is integer. 
Since the output port should be dark in the absence of SGW, 
$\ell_{1}$ and $\ell_{2}$ should 
satisfy
\begin{equation}
\Omega_{\rm L}(\ell_{2}-\ell_{1})=m\pi,
\end{equation}
where $m$ is integer. 
Hence the output in the presence of SGW is given by
\begin{eqnarray}
A_{\rm out}&=&A_{1}+A_{2}
=-\nu_{\rm B}\rho_{\rm B}\rho_{\rm M}A_{\rm L}
e^{i\Omega_{\rm L}(t_{\rm B}-2\ell_{1}-\ell_{\rm E})}
(\Delta A_{1}-\Delta A_{2}), \nonumber \\
&=&\nu_{\rm B}\rho_{\rm B}\rho_{\rm M}A_{\rm L}\Omega_{\rm L}
e^{i\Omega_{\rm L}(t_{\rm B}-2\ell_{1}-\ell_{\rm E})}
\int_{-\infty}^{+\infty}\delta{\tilde \Phi}(\omega)e^{i\omega t_{\rm B}}
H_{\rm FP}(\omega)d\omega,
\end{eqnarray}
where $H_{\rm FP}(\omega)$ is the frequency response function of the 
FP interferometric detector to SGW. 

For simplicity, we focus on the case 
\begin{equation}
\ell_{1}=\ell_{2}=\ell~~~~~{\rm and}~~~~~L_{1}=L_{2}=L.
\end{equation}
Then the response function $H_{\rm FP}(\omega)$ is written as
\begin{eqnarray}
H_{\rm FP}(\omega)&=&-{1\over2\omega}
\Biggl[
\sin\theta\cos\varphi
\biggl\{(2a-b)e^{i\omega(X_{\rm f}-\ell)}
-(a-b)e^{-2i\omega \ell}-a \nonumber \\
&-&{ib\over2\sin(\omega L)}
\left(2e^{i\omega(X_{\rm e}-\ell)}-e^{i\omega(X_{\rm f}-\ell+L)}
-e^{i\omega(X_{\rm f}-\ell-L)}\right) 
\biggr\} \nonumber \\
&-&\sin\theta\sin\varphi
\biggl\{(2a-b)e^{i\omega(Y_{\rm f}-\ell)}
-(a-b)e^{-2i\omega \ell}-a \nonumber \\
&-&{ib\over2\sin(\omega L)}
\left(2e^{i\omega(Y_{\rm e}-\ell)}-e^{i\omega(Y_{\rm f}-\ell+L)}
-e^{i\omega(Y_{\rm f}-\ell-L)}\right) 
\biggr\}
\Biggr],
\end{eqnarray}
where
\begin{eqnarray}
a&\equiv& 1-{\nu_{\rm M}^{2}\over
\rho_{\rm M}(1-\rho_{\rm M})}. \\
b&\equiv& \left({\nu_{\rm M}\over \rho_{\rm M}}\right)^{2}
\left(
{\rho_{\rm M}e^{-2i\omega L}\over 1-\rho_{\rm M}e^{-2i\omega L}}
-{\rho_{\rm M}\over 1-\rho_{\rm M}}
\right).
\end{eqnarray}

\begin{figure}[htbp]
        \centerline{\epsfxsize 13cm \epsfysize 10cm \epsfbox{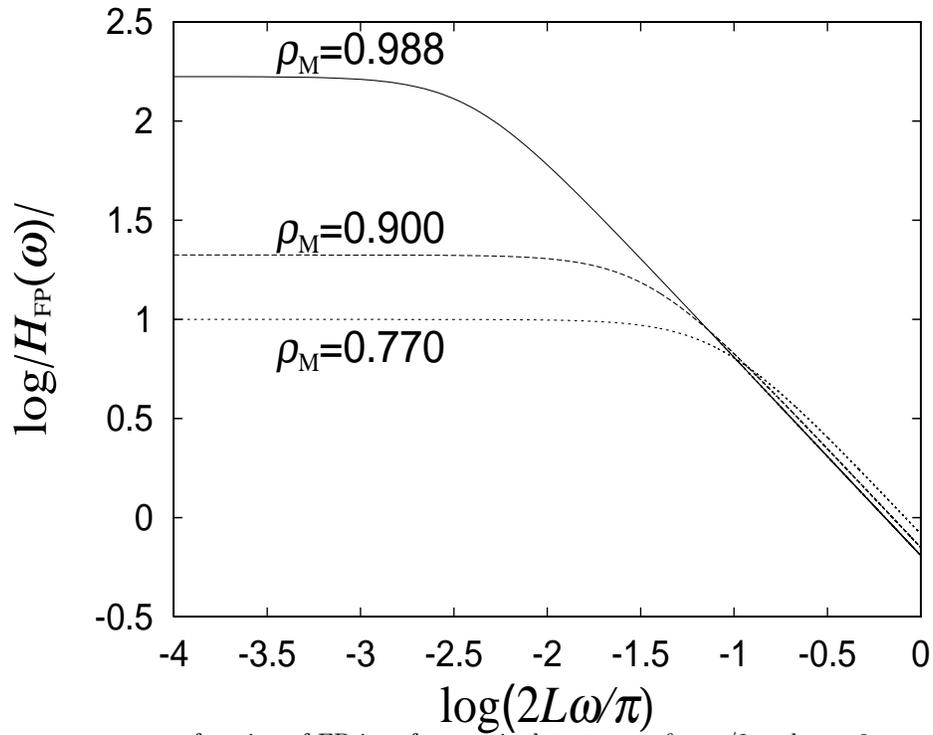}}
        \caption{The frequency response function of FP 
interferometric detectors at $\theta=\pi/2$ and $\varphi=0$ 
as a function of $\omega$. In this figure, we adopt the units in which 
$L$ is unity. The solid, dashed, and dotted lines denotes for 
$\rho_{\rm M}=0.988$, 0.900 and 0.770, respectively. 
For $\rho_{\rm M}=0.770$,  the same sensitivity becomes almost same as that 
for Michelson interferometric detectors. }
        \label{fg:FP-response}
\end{figure}

Due to the same reason as in the case of DL interferometric 
detectors, we assume that $\ell$ is much smaller than $L$. 
Then for $\omega\ell \ll1$, the frequency response function 
$H_{\rm FP}(\omega)$ is derived as 
\begin{equation}
H_{\rm FP}(\omega)\sim
{ibe^{i\omega L}\over 4\omega\sin(\omega L)}H_{\rm M}(\omega;L).
\end{equation}
Thus, FP interferometric detectors also have the same angular
dependence of the frequency response function as Michelson detectors 
of the arm length $L$. 
In the limit $\omega\rightarrow0$, the response 
function $H_{\rm FP}(\omega)$ becomes 
\begin{equation}
\lim_{\omega\rightarrow 0}H_{\rm FP}(\omega)
={i(1+\rho_{\rm M})\over \rho_{\rm M}(1-\rho_{\rm M})}~L
\sin^{2}\theta\cos2\varphi. \label{eq:FP-limit-value}
\end{equation}
In the case of finite $\omega$, 
the absolute value of the 
response function $|H_{\rm FP}(\omega)|$ for $\theta=\pi/2$ and 
$\varphi=0$ becomes
\begin{equation}
|H_{\rm FP}(\omega)|
={(1+\rho_{\rm M})\over \rho_{\rm M}\omega}
\sqrt{1-\cos(2\omega L)\over 2\{1+\rho_{\rm M}^{2}
-2\rho_{\rm M}\cos(2\omega L)\}}. 
\end{equation}
We show $|H_{\rm FP}|$ for various $\rho_{\rm M}$ 
in Fig. \ref{fg:FP-response}. 
Note that $|H_{\rm FP}|$ 
depends on the value of the reflection coefficient $\rho_{\rm M}$ 
of the front mirror sensitively. 
For example, in the case of TAMA300, 
$\rho_{\rm M}$ is equal to $0.988$. 
The interferometric detector such large $\rho_{\rm M}$
is sensitive for $\omega\leq 10^{-2}\pi/(2L)$ for which 
the wave length is much longer than that of the arm length 
of the interferometer. Hence the antenna sensitivity 
pattern of such a FP interferometric detector is almost 
the same as Eq. (\ref{eq:FP-limit-value}). 
This result is consistent with Maggiore and Nicolis \cite{Ref:MN}. 

\section{Summary}

We have rigorously analyzed the sensitivity of 
interferometric detectors to SGW. 
Since we calculate the side-band waves 
of laser rays directly solving the Maxwell field equation, 
we were able to obtain the dependence of the
antenna sensitivity pattern on the frequency of SGW 
without any ambiguity. 
As a result, we found that the antenna sensitivity pattern 
of the simple Michelson 
interferometric detector depends strongly on the frequency of SGW. 
These features are essentially the same as in the case of 
tensor mode of gravitational waves \cite{Ref:T-Mode}. 
LISA is the interferometric detector and will be simply Michelson. 
Thus, the dependence of the 
antenna sensitivity pattern on the wave length of SGW 
has to be taken into account as in the case of the tensor modes. 

We have also found that the dependence of the antenna sensitivity patterns 
of both the Delay-Line and Fabry-Perot interferometers 
on the frequency of SGW is proportional 
to that of the simple Michelson one. This implies that 
if the wave length of SGW is comparable to the 
arm length of the interferometric detector, 
the antenna sensitivity patterns of Delay-Line- and Fabry-Perot 
interferometric 
detectors are different from the result obtained by Maggiore and 
Nicolis who assumed that the amplitude 
of SGW is constant in their calculation. 
However, the effective optical path length of 
Delay-Line- and Fabry-Perot detectors 
will be optimized for SGW 
of the wave length much longer than the arm length. In this case, 
the wave length of SGW is much longer than the arm 
length of the interferometric detector and the amplitude of SGW 
can be considered to be effectively constant for evaluating 
the response, even if the wave length of SGW 
is comparable to the optical length. Therefore, 
the antenna sensitivity 
pattern is almost same as the results obtained by 
Maggiore and Nicolis \cite{Ref:MN}.

\vskip 0.3in
\centerline{ACKNOWLEDGMENTS}
\vskip 0.05in

We would like to thank Bruce Allen for comments.
This work was supported by the 
Grant-in-Aid for Scientific Research (No.05540)
and for Creative Basic Research (No.09NP0801)
from the Japanese Ministry of
Education, Science, Sports and Culture.
MS was supported by JSPS.

\newpage

\begin{figure}
        \centerline{\epsfxsize 10cm \epsfysize 10cm \epsfbox{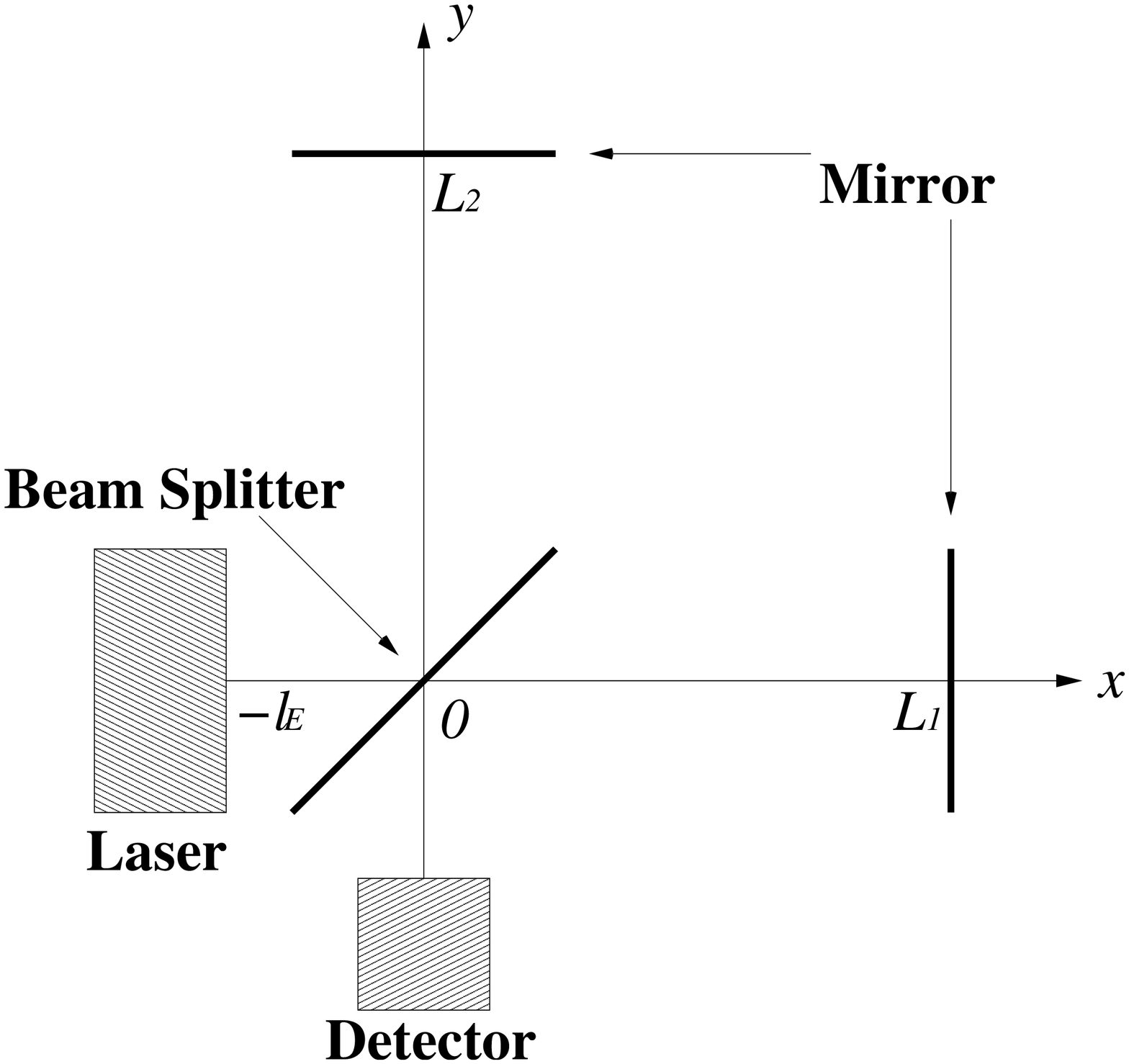}}
~(a)
\end{figure}

\newpage

\begin{figure}
        \centerline{\epsfxsize 13cm \epsfysize 10cm \epsfbox{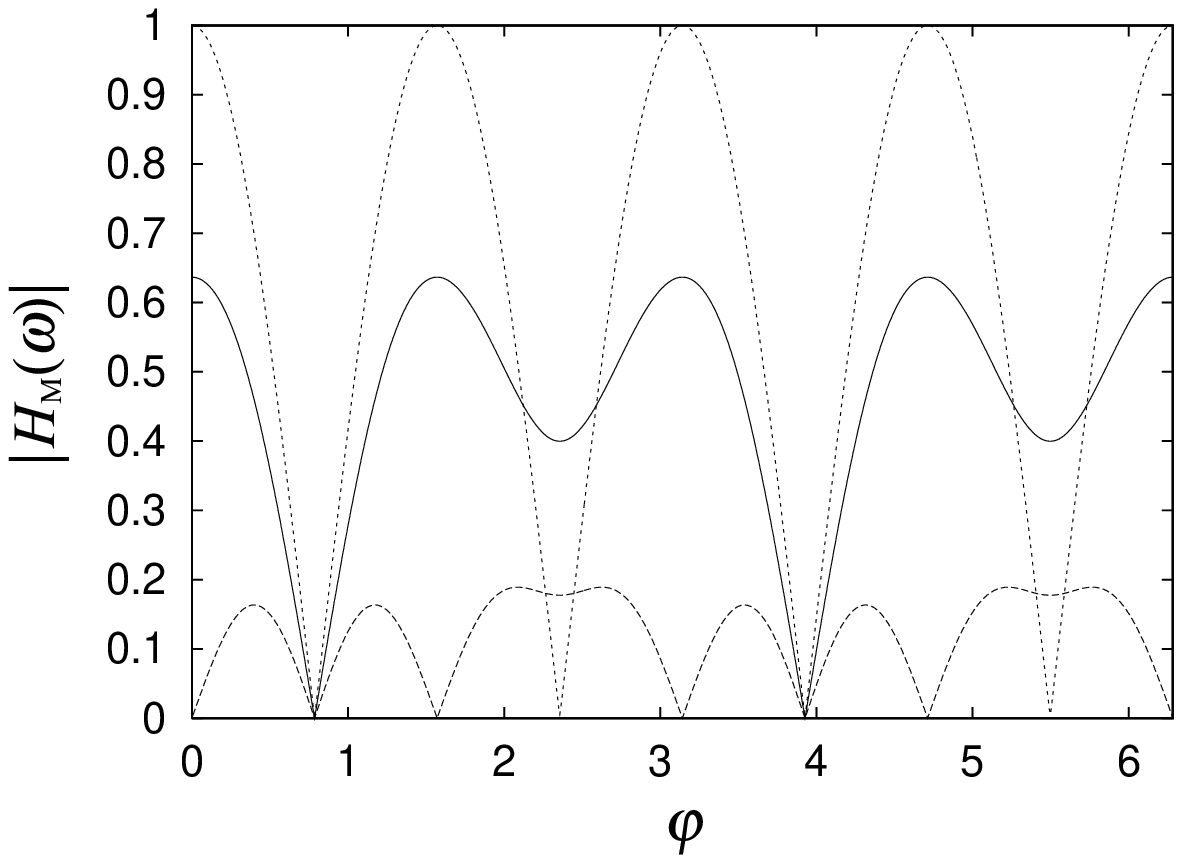}}
~(a)
\end{figure}

\newpage

\begin{figure}
        \centerline{\epsfxsize 10cm \epsfysize 10cm \epsfbox{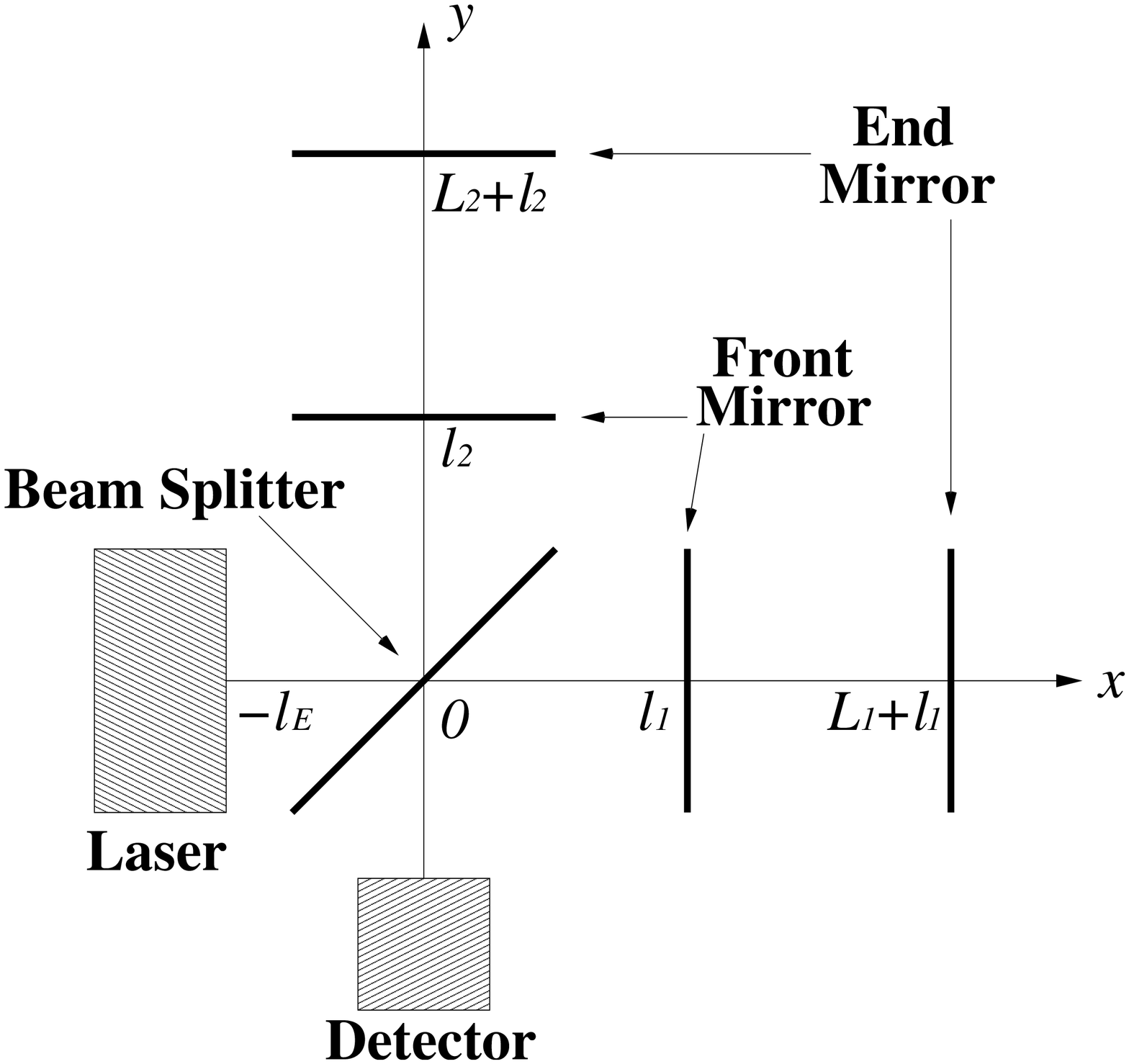}}
~(a)
\end{figure}

\newpage

\end{document}